\documentclass[%
 amsmath,amssymb,
 aps, prl,
 twocolumn,superscriptaddress
]{revtex4-2}

\usepackage{graphicx}
\usepackage{dcolumn}
\usepackage{bm}
\usepackage{color}

\newcommand {\mgu}[1]{\textcolor[rgb]{0,0,0}{#1}}

\begin{document}


\title{Ferroelectric switchable altermagnetism}%

\author{Mingqiang Gu}
\affiliation{Department of Physics and Guangdong Basic
Research Center of Excellence for Quantum Science, Southern University
of Science and Technology, Shenzhen 518055, China}
\author{Yuntian Liu}
\affiliation{Department of Physics and Guangdong Basic
Research Center of Excellence for Quantum Science, Southern University
of Science and Technology, Shenzhen 518055, China}
\author{Haiyuan Zhu}
\affiliation{Department of Physics and Guangdong Basic
Research Center of Excellence for Quantum Science, Southern University
of Science and Technology, Shenzhen 518055, China}
\author{Kunihiro Yananose}
\affiliation{Korea Institute for Advanced Study, Seoul 02455,
Republic of Korea}
\author{Xiaobing Chen}
\affiliation{Department of Physics and Guangdong Basic
Research Center of Excellence for Quantum Science, Southern University
of Science and Technology, Shenzhen 518055, China}
\author{Yongkang Hu}
\affiliation{Department of Physics and Guangdong Basic
Research Center of Excellence for Quantum Science, Southern University
of Science and Technology, Shenzhen 518055, China}
\author{Alessandro Stroppa}
\email{alessandro.stroppa@spin.cnr.it}
\affiliation{CNR-SPIN, c/o Dip.to di Scienze Fisiche e Chimiche - Universit\`{a} degli Studi dell'Aquila - Via Vetoio - 67100 - Coppito (AQ), Italy}
\author{Qihang Liu}
\email{liuqh@sustech.edu.cn}
\affiliation{Department of Physics and Guangdong Basic
Research Center of Excellence for Quantum Science, Southern University
of Science and Technology, Shenzhen 518055, China}
\affiliation{Guangdong Provincial Key Laboratory of Computational Science and Material Design, Southern University of Science and Technology, Shenzhen 518055, China}

\date{\today}

\begin{abstract}
We propose a novel ferroelectric switchable altermagnetism
effect, the reversal of ferroelectric polarization is coupled to the
switching of altermagnetic spin splitting. We demonstrate the design
principles for the ferroelectric altermagnets and the additional
symmetry constraints necessary for switching the altermagnetic spin
splitting through flipping the electric polarization based on the
state-of-the-art spin-group symmetry techniques. 22 ferroelectric
altermagnets are found by screening through the 2001 experimental
reported magnetic structures in the MAGNDATA database and 2 of them are
identified as ferroelectric switchable altermagnets. Using the hybrid
improper ferroelectric material
{[}C(NH\textsubscript{2})\textsubscript{3}{]}Cr(HCOO)\textsubscript{3}
as an example, we show how the altermagnetic spin splitting is tightly
coupled to the ferroelectric polarization, providing an ideal platform
for designing electric-field-controllable multiferroic devices. Finally,
we find that such manipulation of altermagnetism can be detected by
monitoring the physical quantities that are related to the non-vanishing
Berry curvature dipole, such as the linearly polarized photogalvanic
spin current.
\end{abstract}

\maketitle



{\it Introduction - }Multiferroic materials exhibit more than one type of ferroic order
simultaneously, such as ferroelectricity (spontaneous electric
polarization) and magnetism (ferromagnetism or antiferromagnetism)
\cite{RN1}. In particular, magnetoelectrically coupled multiferroic
materials have attracted significant interest due to their potential for
controlling magnetic properties through electric fields, and vice versa,
opening pathways for innovative applications in memory storage, sensors,
and spintronics \cite{RN2,RN3,RN4,RN5}. Multiferroics are broadly categorized into
two types \cite{RN6}. In type-I multiferroics, ferroelectric and magnetic
orders originate from distinct mechanisms \cite{RN7,RN8}, which often results
in weak coupling between these order parameters. In contrast, type-II
multiferroics demonstrate stronger coupling because ferroelectricity
arises directly from magnetic ordering \cite{RN9}. Historically, the design
and manipulation of multiferroicity primarily rely on the interaction
between net magnetic moments and electric polarization \cite{RN10}. For
example, in multiferroic antiferromagnet
Ca\textsubscript{3}Mn\textsubscript{2}O\textsubscript{7}, the
multiferroic coupling stems from the ferroelectric polarization and the
net magnetic moment induced by spin-orbit coupling (SOC), i.e.,
Dzyaloshinskii--Moriya interaction \cite{RN11}. However, the small weak
FM magnetization (0.045 $\mu_{B}$/Mn in theory and 0.0025
$\mu_{B}$/Mn in experiment) \cite{RN12} limits the strength of
magnetoelectric coupling and thus the multiferroic applications.

Recently, altermagnetism has garnered considerable attention \cite{RN_A1,RN13,RN14,RN_A5} 
as a promising avenue for achieving novel spintronic properties.
Altermagnets are a class of collinear antiferromagnetic (AFM) materials
characterized by alternating spin polarization across reciprocal space
due to breaking of certain spatial symmetries, even though they have no
net macroscopic magnetization. The order parameter \emph{S} featuring
altermagnetism can be represented by the energy splitting between the
two spin channels in certain paths of the Brillouin zone
\(\Delta E_{k}^{s} = E_{k}^{\uparrow} - E_{k}^{\downarrow}\). The
spin-split bands of Bloch electrons in altermagnets create the
possibility of designing new types of spintronic devices, e.g.,
spin-filtering magnetic tunnel junctions \cite{RN15,RN16}. Due to the
combination of the advantages of FM and AFM in terms of spin splitting,
altermagnetism is also considered a third type of collinear magnetism
\cite{RN13,RN17}.

In this \emph{Letter}, we introduce the \emph{ferroelectric switchable
altermagnets}, a new class of materials which shows not only the
coexistence of ferroelectricity and altermagnetism, but also the
synergistic coupling of the two order parameters, leading to a reversal
of spin-character in the splitted energy bands by switching the
ferroelectric polarization. As illustrated in Fig. 1(a), in specific
circumstances, the altermagnetic spin degree of freedom is tightly
correlated to the electric polarization, indicating a distinct form of
magnetoelectric coupling. Such a mechanism enables the control of a
nonvolatile spin-split electronic dispersion through the external
electric field, providing an efficient route to achieve robust
all-in-one multiferroic memory devices. For example, Fig. 1(b)
illustrates an electric-field-controllable spin-filtering altermagnetic
tunnel junction device \emph{without the need of switching the N\'eel
vector}. By considering whether the spin-polarized Fermi surface between
the free layer and the fixed layer match or not \cite{RN13}, the conductance $\sigma$ can
be switched between high and low by simply switching the ferroelectric
polarization. By applying the state-of-the-art spin-group symmetry
analyses, we searched through all the MAGNDATA database \cite{RN18} and
screened out 22 \emph{ferroelectric altermagnets}, in which only two
candidates support the synergistic switching of altermagnetic spin
splitting through ferroelectric polarization.

In order to discuss a representative case of the ferroelectric
switchable altermagnets, we perform first-principles calculations on the
metal-organic framework (MOF) system
{[}C(NH\textsubscript{2})\textsubscript{3}{]}Cr(HCOO)\textsubscript{3},
which we will denote as Cr-MOF hereafter. This compound is isostructural
to Cu-MOF found in MAGNDATA and is particularly interesting since it
expands the search for potentially interesting altermagnets to the
extremely tunable and versatile class of hybrid organic-inorganic
perovskites for functional materials design. Cr-MOF has been
experimentally synthesized, and it has been proposed as hybrid improper
ferroelectric \cite{RN19,RN20}. Here, we predict that the altermagnetic spin
splitting in this material is tightly coupled to the ferroelectric
polarization, in both the magnitude and the sign of
\(\Delta E_{k}^{s}\), providing an ideal platform for designing
electric-field-controllable spintronic devices. Finally, we find that
the manipulation of altermagnetism can be monitored by the nonlinear
optical generation of spin current, which originates from the non-zero
Berry curvature dipole.

\begin{figure}
\includegraphics[width=0.75\linewidth]{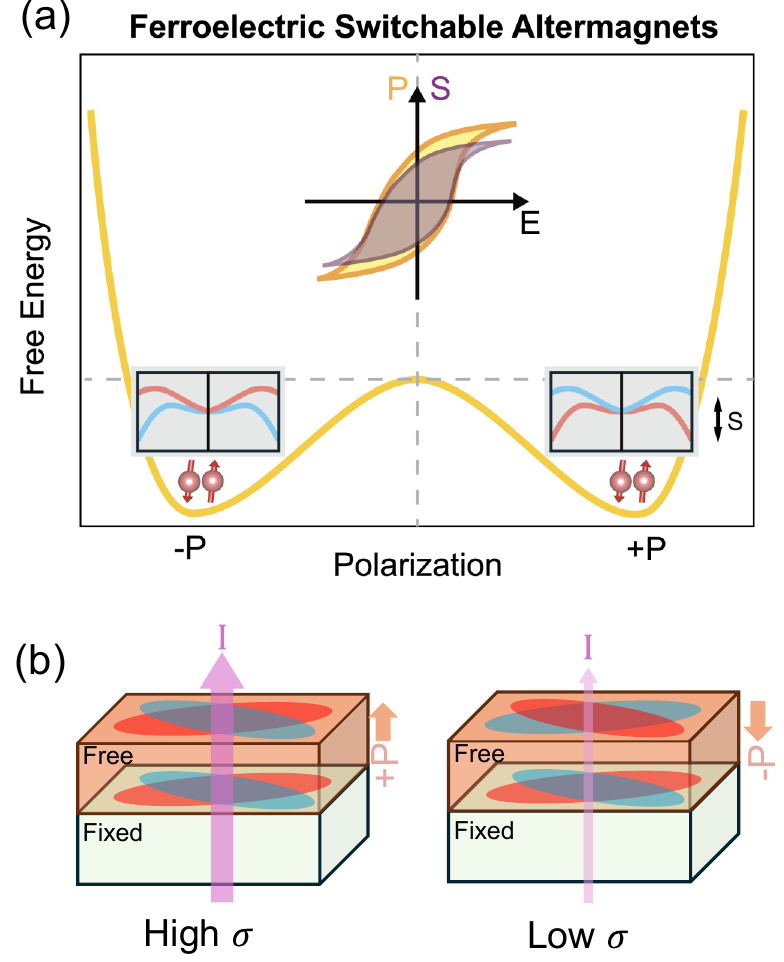}
\caption{\label{fig:fig1}(a) Schematic illustration of ferroelectric switchable
altermagnetism, where the altermagnetic spin splitting \emph{S} strongly
couples to the ferroelectric polarization \emph{P}. (b) A non-volatile
spin filtering tunnel junction device design with ferroelectric
switchable altermagnets. \mgu{The Fermi surfaces of the fixed layer and the free layer are also illustrated. The electric polarization $P$ can be used to control the spin polarization of the free layer, leading to a transition between high- and low-conductance ($\sigma$) states.}}
\end{figure}


{\it Material design principles - } Altermagnetic spin splitting is a non-relativistic property that
manifests without the need for spin-orbit coupling (SOC), where the
symmetry-theoretical framework should adopt spin space group (SSG)
\cite{RN14,RN21,RN22,RN23,RN24,RN25} instead of the commonly used magnetic space group (MSG).
We first consider ferroelectric altermagnets where the two order
parameters \emph{P} and \emph{S} are not necessarily coupled. Two rules
under the framework of SSG emerge: i) For a collinear antiferromagnet
manifesting spin splitting in the momentum space, the symmetries of
joint parity and time-reversal (\(\mathcal{PT}\)) and \(\mathcal{T}\tau\)
with $\tau$ being the fractional lattice translation, should be broken; ii)
For ferroelectric material, the spatial part of the parent space group
(SG) of the SSG should belong to a polar group.

We diagnosed the SSGs of the 2001 experimentally reported magnetic
structures in the MAGNDATA database by using our homemade online program
FINDSPINGROUP \footnote{Https://www.findspingroup.com/}, and found 22 systems that meet the above
criteria. One immediately finds that the filtered materials can be
divided into two classes by examining the MSG of the materials: the MSG
of the first class allows weak spin canting, leading to a non-zero
residual ferromagnetic (FM) moment, while the second class forbids net
magnetic moment in the system. For the first class with \emph{P-M}
coupling, the electronic polarization can interplay with the weak FM
moment, leading to a classical magnetoelectric coupling \cite{RN11}. Note
that SOC plays an essential role (e.g., Dzyaloshinskii-Moriya
interaction) in slightly tilting the spins from the collinear AFM
configuration. For the second class without \emph{P-M} coupling, since
the net magnetic moment is forbidden by MSGs, the flipping of electric
polarization is not supposed to induce any macroscopic effect on the
magnetization. However, both two classes are potential candidates for
ferroelectric switchable altermagnets if the real-space electric
polarization and the momentum-space spin splitting are coupled, even
without SOC.

In general, the coexistence of electric polarization \emph{P} and
altermagnetic spin splitting \(\Delta E_{k}^{S}\) (denoted by \emph{S})
does not guarantee their coupling. Therefore, we next focus on the
search of symmetry operations that synergistically switch \emph{P} and
\emph{S}, and if these operations correspond to reasonable flipping
paths under an external electric field, which does not change the N\'eel
order. For example, considering an initial state \(\psi_{k}^{i}(P,S)\),
one can always flip both electric and spin polarization by applying a
joint operation \(\mathcal{PT}\) to the system to get a reversed state
\(\mathcal{PT}\psi_{k}^{i}(P,S) = \psi_{k}^{f}( - P, - S)\). However, in
the transition path connecting \(\psi_{k}^{f}( - P, - S)\) and
\(\psi_{k}^{i}(P,S)\) phases, the N\'eel vector may also be reversed,
indicating a transition path that cannot be achieved by electric field
only. By going through the list in TABLE SI in Supplementary Materials,
we found that the sign of \(\Delta E_{k}^{S}\) for most candidates
remains unchanged during the polarization reversal, as exemplified by
PbNiO\textsubscript{3} in Supplementary Materials Sec. III. Interestingly, there
are two hybrid improper ferroelectric materials,
Ca\textsubscript{3}Mn\textsubscript{2}O\textsubscript{7} and
{[}C(NH\textsubscript{2})\textsubscript{3}{]}Cr(HCOO)\textsubscript{3},
as prototypical candidates of ferroelectric switchable altermagnets.

\begin{figure}
    \centering
    \includegraphics[width=1\linewidth]{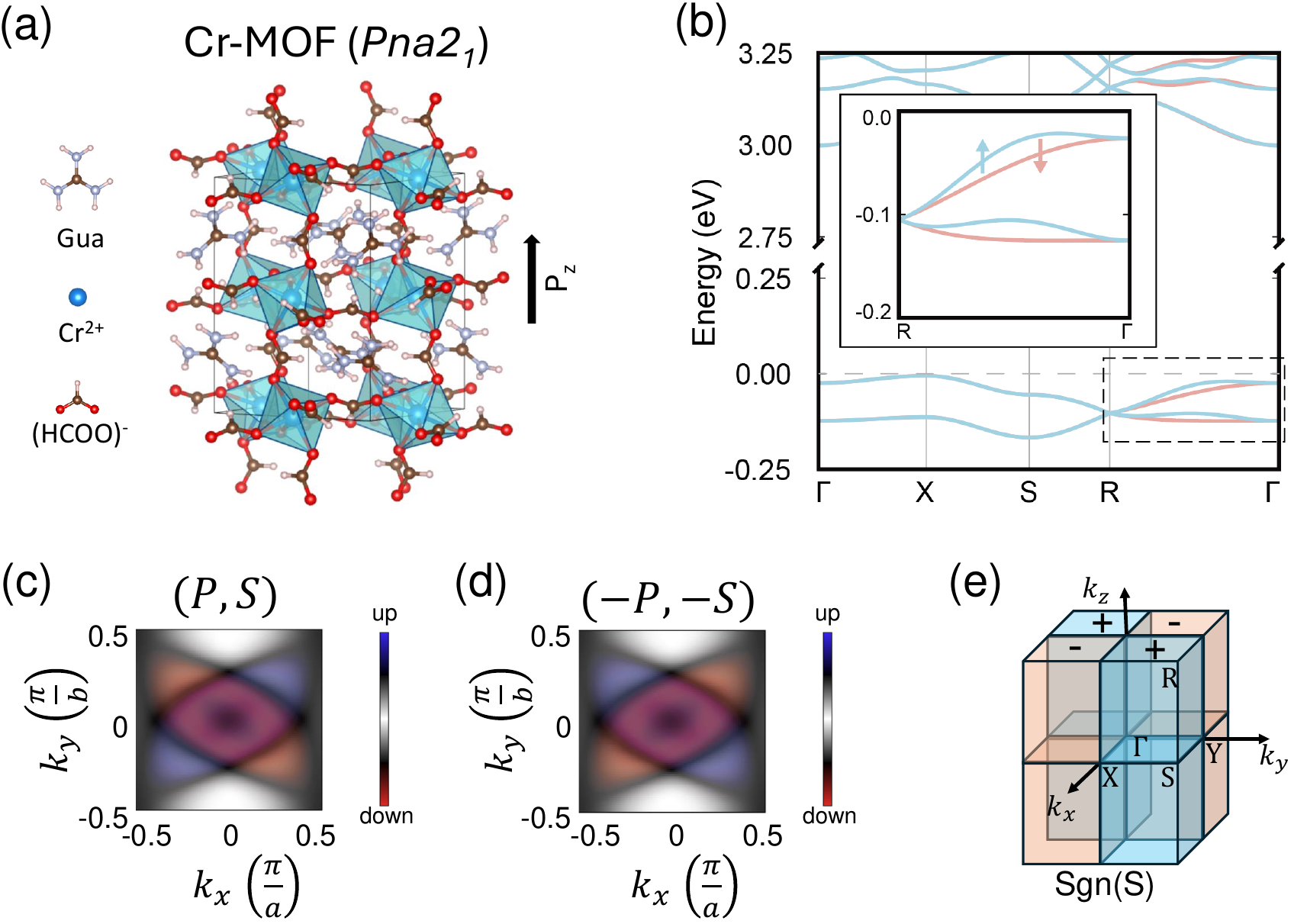}
    \caption{(a) Crystal structure of Cr-MOF. (b) Spin-split band
structure of Cr-MOF. (c) Spin-resolved isoenergy contour ($E=-0.05$ eV) for the
\((P,S)\) state at the \(k_{z} = 0.25\) plane, blue and red denote the
spectral weight for spin up and spin down, respectively. (d) Same as
(c), but for the \(( - P, - S)\ \)state. (e) The sign of
\(\Delta E_{k}^{S}\) for the VBM in reciprocal space.}
    \label{fig:fig2}
\end{figure}

{\it Altermagnetism in hybrid improper ferroelectric Cr-MOF -}
Hybrid improper ferroelectricity arises from a combination of two or
more lattice modes that individually preserve inversion symmetry but
collectively breaks it, resulting in a spontaneous polarization. Several
candidate materials from Class 1 in TABLE SI belong to this category,
including Ca\textsubscript{3}Mn\textsubscript{2}O\textsubscript{7} (SG
\(Cmc2_{1}\)) \cite{RN26} and the hybrid organic-inorganic perovskite
{[}C(NH\textsubscript{2})\textsubscript{3}{]}M(HCOO)\textsubscript{3}
(SG \(Pna2_{1}\)) with M being transition-metal elements Cu or Cr. Below
we use Cr-MOF (see Fig. 2(a)) to illustrate the design principles of
ferroelectric switchable altermagnets for two reasons: (a) it exhibits
the largest spin splitting near the Fermi level among this material
family, and (b) various AFM configurations can be realized in Cr-MOF
through compressive strain, offering an additional degree of freedom for
phase manipulation \cite{RN27}. Moreover, the compositional tunability in
this family offers more flexibility in designing materials for
spin-splitting magnitude optimization. While A-, C- and G-type AFM
Cr-MOF are all altermagnets, here we focus on the C-type configuration,
as it manifests the largest spin splitting near the Fermi level. Further
discussions on other MOFs and AFM configurations can be found in the
Supplementary Materials Sec. IV.

The spin-polarized band structure obtained from density functional
theory (DFT, see Supplementary Materials Sec. I for details) calculations is shown in Fig. 2(b). The band gap is about 3
eV, ensuring a good insulating background that is favorable for
ferroelectrics. Notably, a large spin splitting is observed at the
valence band maximum (VBM), with the largest energy difference being
approximately 20 meV. Such spin splitting occurs at the relatively
low-symmetry points in reciprocal space, such as along the
\(\Gamma(0,0,0) \rightarrow R(0.5,0.5,0.5)\) path. The spin-polarized
isoenergy contour at \(E = - 0.05\) eV is shown in Fig. 2(c), showing a
deviated feature compared with the so-called ``d-wave'' altermagnetism
because of the absence of four-fold symmetry. Furthermore, the sign of
\(\Delta E_{k}^{S}\), a well as the spin polarization of a given
spin-polarized band, are odd under mirror operations
\{1\textbar\textbar{}\emph{m}\textsubscript{100}\} and
\{1\textbar\textbar{}\emph{m}\textsubscript{010}\}, and even under
\{1\textbar\textbar{}\emph{m}\textsubscript{001}\}, as illustrated in
Fig. 2(e).

\begin{figure*}
    \centering
    \includegraphics[width=\linewidth]{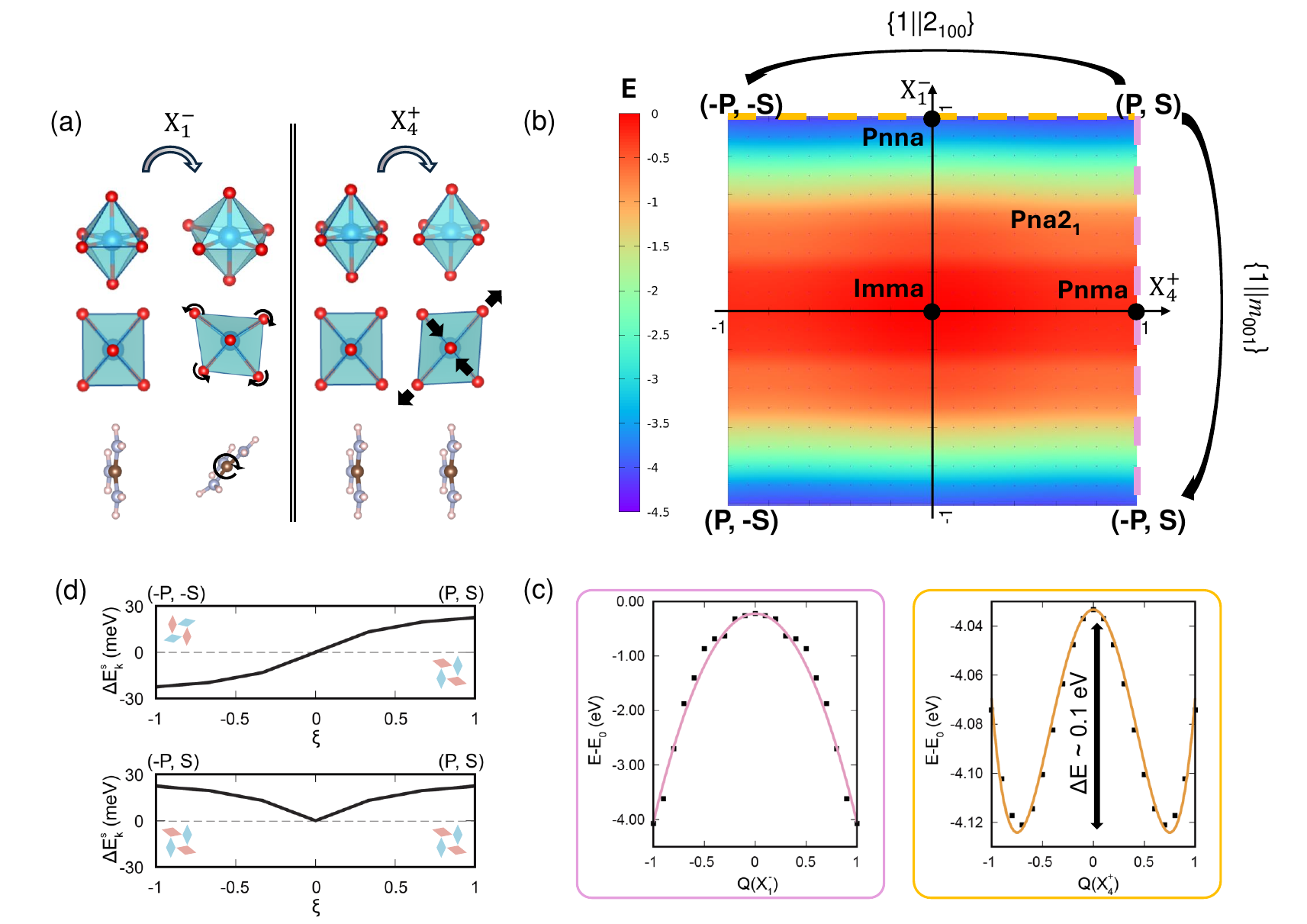}
    \caption{(a) The hybrid improper coupled zone-boundary
distortion modes in Cr-MOF. (b) Energy landscape for the coupling of the
\(X_{1}^{-}\) and \(X_{4}^{+}\) modes, where different transition paths
are marked by yellow and purple dashed lines. (c) Energy profiles along
the two paths denoted in (b). (d) Spin splitting as a function of the
interpolation structural parameter $\xi$. The insets in (d) denote the
Q\textsubscript{2} JT pattern at the endpoints, with the pink and blue
diamonds denote the octahedra centered with the spin-up and spin-down
Cr, respectively. The elongated octahedra form an antiferro-distortive
JT distortion pattern associated to the orbital-ordering which
``pseudo-rotate'' from the two endpoints \((P,S)\) and \(( - P, - S)\).}
    \label{fig:fig3}
\end{figure*}

{\it Synergistically switch of electric and spin polarizations -}
Here, a state with the coexistence of polarization \(P\) (\(P//z\)) and
altermagnetism \(S\) is denoted as the \((P,S)\) state. The SSG
operations that map \(P\rightarrow-P\) and \(S\rightarrow-S\) are analyzed and listed in Fig. S3.
While the SSG operations connecting \(P\) to \(-P\) state can be one
of the following: \{1\textbar\textbar-1\},
\{1\textbar\textbar2\textsubscript{100}\},
\{1\textbar\textbar2\textsubscript{010}\}, \{-1\textbar\textbar-1\},
\{-1\textbar\textbar2\textsubscript{100}\}, and
\{-1\textbar\textbar2\textsubscript{010}\}, only three of them map the
\((P,S)\)state to the \((-P,-S)\) state, i.e.,
\{1\textbar\textbar2\textsubscript{100}\},
\{1\textbar\textbar2\textsubscript{010}\}, and
\{-1\textbar\textbar-1\}. In addition, whether the \((-P,-S)\) state
can be experimentally realized upon the application of an electric
field, however, depends on the energy barrier between the two phases,
which is ultimately governed by the internal atomic displacements during
the phase transition.

The search for the reversed polarization state can also be achieved by
considering the mode decomposition of hybrid improper ferroelectricity.
The electric polarization of Cr-MOF along the \emph{z} direction is
activated by two non-polar lattice distortion modes through a trilinear
coupling \(F = \gamma Q_{X_{1}^{-}}Q_{X_{4}^{+}}Q_{P}\), where
\(Q_{X_{1}^{-}}\), \(Q_{X_{4}^{+}}\) and \(Q_{P}\) represent the
amplitudes of the two zone-boundary non-polar distortion modes and the
polar distortion mode, respectively, relative to the high symmetry
\emph{Imma} phase \cite{RN19}. Reversing either of the zone-boundary modes
flips the polarization \(P \rightarrow - P\), while, reversing both
preserves \(P\). As shown in Fig. 3(a), the \(X_{1}^{-}\) mode consists
of two major distortions: a ``scissor''-like distortion for the
equatorial Cr-O bonds and a rotation of the guanidinium molecule by
about 42°. The \(X_{4}^{+}\) mode corresponds primarily to a
Q\textsubscript{2} Jahn-Teller (JT) distortion of the Cr octahedra, with
minimal impact on the organic molecules. Starting from the \emph{Imma}
structure, imposing only the \(X_{1}^{-}\) (\(X_{4}^{+}\)) mode results
in a centrosymmetric \emph{Pnna} (\emph{Pnma}) structure. The
combination of both \(X_{1}^{-}\) and \(X_{4}^{+}\) modes leads to the
\emph{Pna2\textsubscript{1}} structure, a polar SG that induces the
softening of the polar distortion. Since the polar distortion is a
secondary effect arising from the coupling of the zone-boundary modes,
the interplay between these two modes fully determines the final
structure and its related properties in this system. For simplicity, and
without loss of generality, we will use these two unstable modes
(\(X_{1}^{-}\) and \(X_{4}^{+}\)) as variables in the discussion below.

Fig. 3(b) shows the energy landscape of the coupled zone-boundary modes
relative to the \emph{Imma} phase. Starting from the energetically
favored \((P,S)\) ground state, if the amplitude of \(X_{1}^{-}\) mode
is continuously reduced and eventually reversed its sign, the structure
transforms to a \(( - P,S)\) state. This transition path, indicated by
the purple line in Fig. 3(b), passes through a centrosymmetric
\emph{Pnma} phase. The \(( - P,S)\) state and the original \((P,S)\)
state are related by \{1\textbar\textbar{}\emph{m}\textsubscript{001}\}
SSG operation. The energy barrier is about 4 eV per unit cell, as shown
in the left panel of Fig. 3(c). Such a high energy barrier is attributed
to the large atomic distortion amplitude ($\sim$ 6.3 Å) along
this path, primarily due to the rotation of the guanidinium molecule. In
sharp contrast, when the \(X_{4}^{+}\) mode is continuously tuned to the
reversed amplitude, the transition occurs via a \emph{Pnna} reference
phase (orange line in Fig. 3(b)). This path leads to the \(( - P, - S)\)
state, connected by \{1\textbar\textbar2\textsubscript{100}\} SSG
operation, with a significantly lower energy barrier of approximately
0.1 eV (the right panel of Fig. 3(c)), nearly 40-times smaller than
that for the \emph{Pnma} path. Thus, the
\((P,S) \rightarrow ( - P, - S)\) structural transition through the
\emph{Pnna} path is energetically favored and represents a possible
realistic transition pathway when polar direction is flipped by an
external electric field.

The spin reversal of the isoenergy contour for the \(( - P, - S)\) state shown in
Fig. 2(d) strongly supports the possibility to control altermagnetic
spin splitting by flipping the ferroelectric polarization in this
system. To quantify this transition, we define a normalized
interpolating structural parameter \(\xi\), where \(\xi = 1\) represents
the \((P,S)\) structure and \(\xi = - 1\) corresponds to either the
\(( - P,S)\) or \(( - P, - S)\) structure. The evolution of
\(\Delta E_{k}^{s}\) during the structural transition is plotted in Fig.
3(d). We found that \(\Delta E_{k}^{s}\) changes monotonically during
the \((P,S) \rightarrow ( - P, - S)\) transition, while in the
\((P,S) \rightarrow ( - P,S)\) transition it reduces to zero at the
\emph{Pnma} phase and then regresses to its original value. Such
behavior arises because \(\Delta E_{k}^{s}\) is odd under
\{1\textbar\textbar2\textsubscript{100}\} but even under
\{1\textbar\textbar{}\emph{m}\textsubscript{001}\}, consistent with the
symmetry presented in Fig. 2(e). Furthermore, the microscopic mechanism
of the reversal of the altermagnetic spin splitting by reversing the
\(X_{4}^{+}\) mode can be understood by the change of the
Q\textsubscript{2} JT distortion pattern, as illustrated in the insets
of Fig. 3(d). Since the JT distortion is known to couple with the
orbital ordering in oxide perovskites \cite{RN28} and MOF systems \cite{RN29},
our results provide a concrete example of how altermagnetism can be
entangled with orbital ordering \cite{RN30}, offering new insights into the
control of spin splitting through orbital as well as lattice degrees of
freedom.

\mgu{
To illustrate the electric-field-controllable altermagnetic tunnel junction device, we consider a prototypical homojunction system proposed in Fig. 1(b), where both the fixed and free layers are composed of Cr-MOF. We evaluate the giant magnetoresistance (GMR) by the ratio between the DFT-calculated longitudinal conductivity of the spin-up and spin-down channels along the [110] direction \cite{RN_A7}. Based on a 184-atom $\sqrt{2}\times\sqrt{2}$ supercell, our results suggest that the maximum GMR rate for this system could reach approximately 53\% (see Fig. S5 and Supplementary Materials Sec. VI for details), validating the potential of ferroelectric switchable altermagnetism for non-volatile spintronic devices.
}


{\it Detection of the switch of altermagnetic spin splitting -}
The direct experimental detection of the AFM-induced spin splitting
typically requires advanced techniques such as spin-resolved
angle-resolved photoemission spectroscopy \cite{RN31,RN32}. Since the N\'eel
vector remains unchanged through \((P,S) \rightarrow ( - P, - S)\)
transition, we propose a way to detect the switch of altermagnetic spin
splitting through the linear-polarized photogalvanic spin current effect
(spin LPGE), which directly couples to the spin polarization through
nonlinear optical response. By using the SSG symmetry analysis
implemented in FINDSPINGROUP, we find that six non-zero Berry curvature
dipole (BCD) tensor elements
(\(D_{xxz},\ D_{xzx},D_{zxx},D_{yyz},D_{yzy},D_{zyy}\)) are allowed
without SOC in Cr-MOF. All the other tensor elements, such as the
quantum metric dipole and inverse mass dipole \cite{RN33}, are forbidden.
The non-zero BCD can induce a nonlinear Hall current in transport
measurement, as well as the spin LPGE from an optical perspective. Due
to the broken spatial inversion symmetry, a shift current is expected
under linear-polarized light excitation \cite{RN34}. In Cr-MOF system, both
charge and spin currents can be generated: the charge current reflects
the direction of the electric polarization \(P\), while the spin current
serves as a signature of the difference between the two spin channels,
i.e. the spin splitting.

The spin LPGE conductivity component
\(\sigma_{\uparrow - \downarrow}^{zyy}\) for the \((P,S)\) and
\(( - P, - S)\) states is shown in Fig. 4. A non-zero dc current begins
to emerge when the photon energy exceeds the band gap, reaching the
first peak at about 3.2 eV with a value of about 0.23 \(\mu A/V^{2}\).
The second peak, near 4 eV, is attributed to the spin splitting at the
higher conduction bands. During the \((P,S) \rightarrow ( - P, - S)\)
structure transition, the sign of the spin LPGE current reverses.
Therefore, the spin LPGE current can serve as a signature of the
reversal of the altermagnetic spin splitting.

\begin{figure}
    \centering
    \includegraphics[width=1\linewidth]{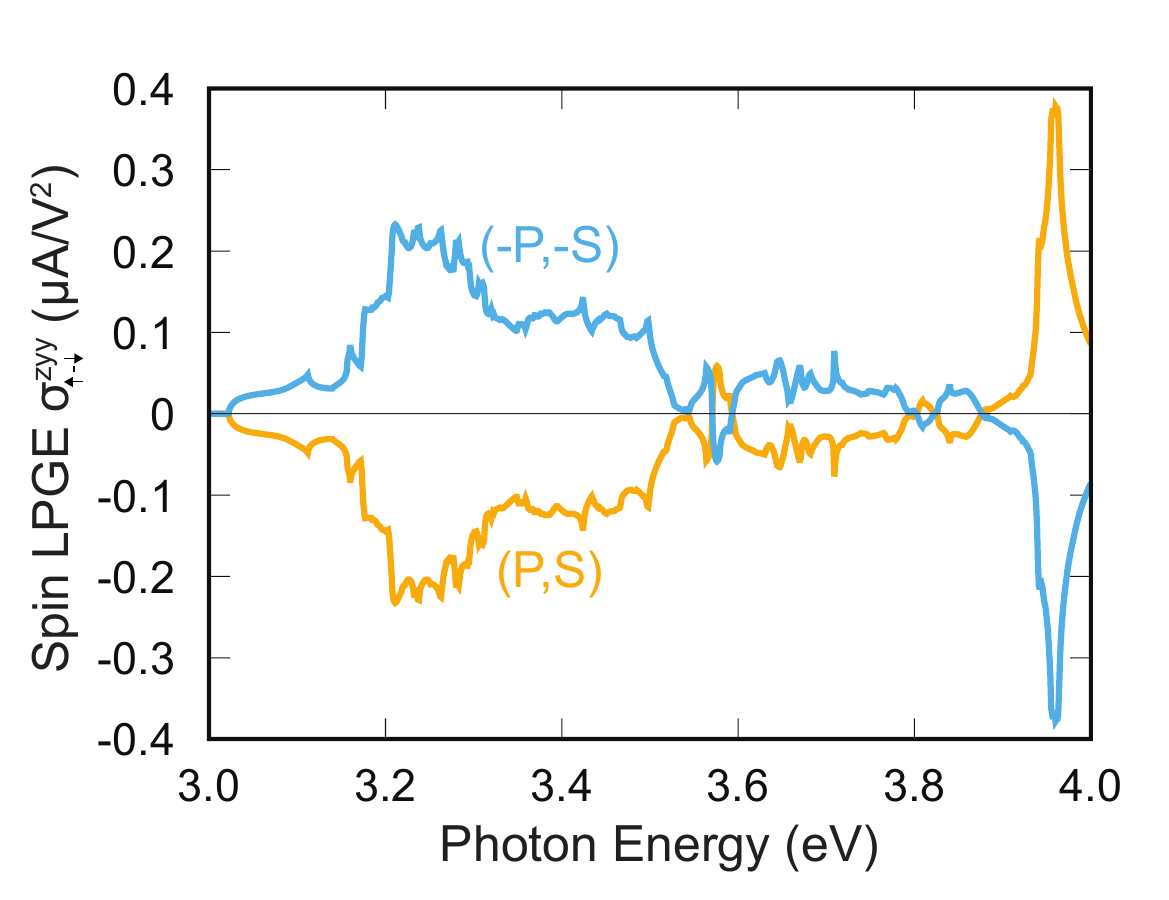}
    \caption{The spin LPGE conductivity component
\(\sigma_{\uparrow - \downarrow}^{zyy}\) for the \((P,S)\) and
\(( - P, - S)\) states.}
    \label{fig:fig4}
\end{figure}

{\it Conclusion and discussion -}
Recently, the gate-field control of different spin channels in layered
altermagnets through spin-valley-layer locking effect has been proposed
\cite{RN35, RN_A2}, highlighting the importance for the manipulation of spin
degree of freedom in altermagnets. However, such a non-versatile
strategy requires the constant application of an external electric
field, which is not favorable for storage devices unless further
integration with other materials. Sliding
ferroelectricity \cite{RN36} or 
\mgu{
phase transition between ferroelectric and antiferroelectric states \cite{RN_A6} 
}
can also be used to tune the \(\mathcal{PT}\) symmetry
in the two dimensional van der Waals systems,
which eventually control the emergence of the altermagnetic state. Considering that the hybrid
improper ferroelectric materials can be artificially designed by
chemistry or strain engineering \cite{RN37}, our study provides a wide
configuration space for designing versatile ferroelectric switchable
altermagnets.

Our approach provides a novel method for controlling spin properties
through ferroelectric polarization, which differs from traditional
magnetoelectric multiferroics where ferroelectricity is used to control
the net magnetic moment. 
\mgu{
For the Class 1 candidate
Ca\textsubscript{3}Mn\textsubscript{2}O\textsubscript{7}, the energy splitting ($\sim$ 100 meV) is even larger than that of Cr-MOF.
Spin canting is allowed in this system, leading to a weak net ferromagnetic moment $M$. 
Our symmetry analysis shows that the order parameter for spin splitting $S$ follows the same transformation rules as that for $M$ \cite{RN11}, which can be detected by the anomalous Hall effect.
Therefore, an additional order parameter $M$ is added to the coupled degrees of freedom, i.e., $(P,M,S)$, offering multiple tuning knobs for material manipulation and providing enhanced versatility in device design (see Supplementary Materials Sec. V for details).
}
In contrast, for the Class 2 candidates with no net magnetic moment, the most traditional magnetic detection
method such as magneto-optical Kerr effect \cite{RN38, RN_A5} could be
effective, our approach still offers a viable pathway to achieve
magnetoelectric coupling by correlating the electric polarization with
spin splitting in reciprocal space. To conclude, ferroelectric
switchable altermagnetism is a new type of magnetoelectric coupling
beyond what is achievable in traditional multiferroics based solely on
the spin-orbit interaction, paving a new functional materials platform
for designing all-electric-controlled spintronic devices.

\mgu{ 
{\it Note added -} Recently, we became aware of a related study on Ca\textsubscript{3}Mn\textsubscript{2}O\textsubscript{7} for the ``altermagnetoelectric effect'' \cite{RN_A3}, a concept similar to the ferroelectric switchable altermagnetism proposed in our work.}

\begin{acknowledgments}
A.S. thanks useful discussions with P. Radaelli (Oxford University), P. Barone (CNR-SPIN),  Sang-Wook Cheong (Rutgers University) and Xinfeng Chen (Xi’an Jiaotong University). 
This work was supported by the National Key R\&D Program of China under
Grant No. 2019YFA0704900, the National Natural Science Foundation of
China under Grant No. 12274194 and 12474232, Guangdong Provincial Key Laboratory for
Computational Science and Material Design under Grant No.
2019B030301001, Shenzhen Science and Technology Program (Grants No.
RCJC20221008092722009 and No. 20231117091158001), the Innovative Team of
General Higher Educational Institutes in Guangdong Province (Grant No.
2020KCXTD001), Guangdong Provincial Quantum Science Strategic Initiative
under Grant No. GDZX2401002, the Science Technology and Innovation
Commission of Shenzhen Municipality (JCYJ20210324104812034), Natural
Science Foundation of Guangdong Province (2021A1515110389) and Center
for Computational Science and Engineering of Southern University of
Science and Technology. This work has been funded by the European Union - NextGenerationEU, Mission 4, Component 1, under the Italian Ministry of University and Research (MUR) National Innovation Ecosystem grant ECS00000041-VITALITY-CUP B43C22000470005. We also acknowledge the support by bilateral
agreement for scientific collaboration CNR and NSFC (China),
``Ferroelectric and chiral hybrid organic inorganic perovskites'', for
the years 2024-2025 (CUP B53C2300716000). K.Y. was supported by
an individual grant (No. CG092501) at Korea Institute for Advanced
Study.
\end{acknowledgments}

\appendix

%

\bibliography{FSA_bib}

\end{document}